\documentclass[aps,pre,twocolumn,superscriptaddress,citeautoscript,amsmath,amssymb]{revtex4}
\usepackage{array}
\usepackage{graphicx}
\usepackage{dcolumn}
\usepackage{bm}
\usepackage{times}

\begin{document}

\preprint{}

\title{Conductance of Ferro- and Antiferro-magnetic single-atom contacts: A first-principles study}

\author{Zhi-Yun Tan}
\affiliation{Department of Physics, Zunyi Normal College, Zunyi, People's Republic of China.}
\affiliation{Department of Physics,  Shanghai Normal University,
100 Guilin Road, Shanghai  200232, People's Republic of China.}

\author{Xiao-long Zheng}
\affiliation{Department of Physics, Shanghai Normal University, 100 Guilin Road, Shanghai
200232, People's Republic of China.}

\author{Xiang Ye}%
\affiliation{Department of Physics, Shanghai Normal University, 100 Guilin Road,
Shanghai 200232, People's Republic of China.}

\author{Yi-qun Xie*}
\email{yqxie@shnu.edu.cn}
\affiliation{Department of Physics, Shanghai Normal University, 100 Guilin Road, Shanghai
200232, People's Republic of China.}
\affiliation{Beijing Computational Science Research Center, 3
Heqing Road, Beijing 100084, People's Republic of China}

\author{San-huang Ke*}
\email{shke@tongji.edu.cn}
\affiliation{Key Laboratory of Advanced Microstructured Materials, MOE, Department of Physics, Tongji University,
1239 Siping Road, Shanghai 200092, People's Republic of China}
\affiliation{Beijing Computational Science Research Center, 3
Heqing Road, Beijing 100084, People's Republic of China}

\date{\today}

\begin{abstract}
We present a first-principles study on the spin denpendent conductance of five
single-atom magnetic junctions consisting of a magnetic tip and an adatom
adsorbed on a magnetic surface, i.e., the Co-Co/Co(001) and Ni-X/Ni(001) (X=Fe,
Co, Ni, Cu) junctions.
When their spin configuration changes from ferromagnetism  to
anti-ferromagnetism, the spin-up conductance increases while the spin-down one
decreases. For the junctions with a magnetic adatom, there is nearly no spin
valve effect as the decreased spin-down conductance counteracts the increased
spin-up one. For the junction with a nonmagnetic adatom (Ni-Cu/Ni(001)),
a spin valve effect is obtained with a variation of $22\%$ in the total
conductance. In addition, the change in spin configuration enhances
the spin filter effect for the Ni-Fe/Ni(001) junction but suppresses it
for the other junctions.

\end{abstract}

\maketitle

\section{Introduction}
Single-atom magnetic junctions (SAMJs) may serve as basic components for
future electronic nanodevices. Therefore, the understanding of their electric transport
property is of the fundamental interest, especially for their potential applications in
spintronics\cite{Spintronics}. So far, many experimental and theoretical studies
have been devoted to quantify  the conductance of SAMJs $^{2-21}$. For example, a number
of studies were focused on investigating and understanding of the conductance of half conductance quantum, i.e.,
$e^2/h$,  where $e$ is the proton charge and $h$ is the Planck's
constant\cite{Ugarte2,stm2,Nielsen,ElChemNi,CuHalfQuantum,Untiedt,Tao,xie}. It
has been found that the spin-dependent conductance of a SAMJ is affected
significantly by its contact atomic structure \cite{TB3Ni,Lazo} and its spin
configuration \cite{Bagrets,stm3,DftNi1}, as well as by the magnetization direction
relative to the crystallographic axes\cite{Neel1}. The change of the
spin configuration from ferromagnetism (FM) to anti-ferromagnetism (AFM)
can cause a change in the total conductance and a spin valve effect which is a
key topic in spintronics. Recently, a voltage-dependent spin valve effect
was observed with a conductance variation of $40\%$ for a SAMJ comprising a
Cr-covered tip and a Co or Cr adatom on Fe nanoscale islands formed on a W(110)
substrate\cite{stm3}. The variation of the total conductance
was found to result mainly from the change of the spin-up conductance.
For other SAMJs with different species and spin configurations, it is still an
open and interesting problem whether this spin valve behavior
also exists and how it is affected by the majority and minority electrons.

Here we present a first-principles investigation to show that there is no spin
valve effect for several SAMJs with a magnetic adatom under zero bias
when the spin configuration changes from FM to AFM.  A
spin valve effect with a large change in the total conductance is however
observed for the junction with a nonmagnetic adatom. These results can be
attributed to the different behavior of the majority and minority electrons
caused by the change of spin configuration.

\section{Theoretical Models and  Methods}

\begin{figure}[tb]
\center
\includegraphics[width=8.5cm]{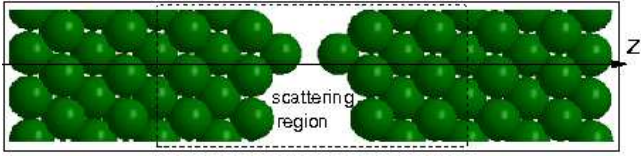} 
\caption{Atomic model for the transport calculations of the SAMJs. 
The transport is along the $z$ direction. }
\label{fig1}
\end{figure}

In this work, a SAMJ is modeled by a tip-adatom/surface junction. Five SAMJs
having the similar atomic geometry are considered, i.e., the Co-Co/Co(001) and Ni-X/Ni(001)
junctions, where X denotes a Fe, Co, Ni, or Cu adatom, respectively. The single
adatom is adsorbed on the hollow site of the $fcc$ (001) surface represented by
a four-layer slab, with each layer containing $3\times3$ atoms as displayed in
Fig.\ref{fig1}. The tip is
modeled by a single apex atom adsorbed on a four-layer (001) slab. The
tip-apex atom is placed above the adatom in the $z$ direction.  In transport
calculations, these atoms construct the scattering region, and four additional
(001) layers are added at the two ends of the scattering region, respectively,
to mimic the left and right electrodes (leads). This kind of structure model has
been proven to be reasonable in describing the spin transport properties of SAMJs in
a previous work\cite{xie}.

The atomic structure of the scattering region is optimized by the VASP code \cite{vasp}.
The two bottom layers of the surface and the top layer of the tip are fixed during the structure
optimization while the other atoms are fully
relaxed until the maximum force is smaller than 0.01eV/\AA. Projector augmented-wave
method \cite{paw} is used for the wave function expansion with an
energy cutoff of 400eV. The PW91 version \cite{PW91} of the generalized gradient
approximation (GGA) is adopted for the electron exchange and correlation.
The Brillouin zone is sampled with a $4\times4\times1$  grid of the Monkhorst-Pack k points \cite{MK}.

For the quantum transport calculation, we adopt the first-principles nonequilibrium Green's
function (NEGF) approach \cite{ke2004,datta95} which combines the
NEGF formula for transport with {\it ab initio} density functional theory (DFT) calculation for electronic
structure. A numerical double zeta plus polarization basis set (DZP) is used for
the wavefunction expansion.
Other computational details are the same as adopted in our previous work\cite{xie}.
The spin-polarization ratio (SPR) at Fermi
energy is defined as $P=(T_\uparrow-T_\downarrow)/(T_\uparrow+T_\downarrow)$, where
$T_\uparrow$ and $T_\downarrow$ denote the transmission coefficient of the
majority and minority spin, respectively. The conductance is scaled in units of $e^2 /h$.

\section{Results and Discussion}

\begin{figure}
\center
\includegraphics[width=8.5cm]{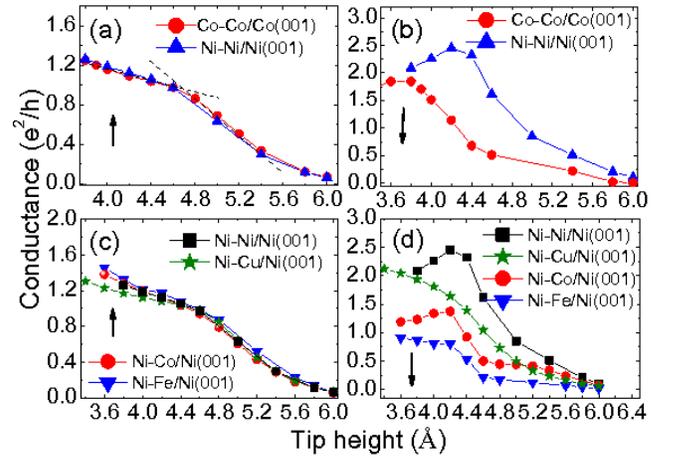}
\caption{Spin-dependent conductance of the SAMJs as a function of the tip height.
The up and down arrows indicate the spin-up and spin-down components, respectively. }
\label{fig2}
\end{figure}

We first calculate the conductance of the SAMJs in FM configuration
as a function of the tip height (the distance between the tip-apex
atom and the surface atom before relaxation) and plot the result in
Fig.\ref{fig2}. In Fig.\ref{fig2}(a), one can see clearly that as the
tip height decreases the conductance increases and shows a faster change in the
transition region, e.g, from 5.4 to 4.8\AA\, and then increases slowly in the contact
region, e.g, below 4.6\AA. The conductance data in the transition and contact regions
can be approximated by two straight lines, and their intersection point defines
the contact conductance\cite{stm2,stm3}. According to this definition, we
obtain the contact conductances of these SAMJs and find that the spin-up
conductances of all the junctions are very close to each other, while the
spin-down conductance varies largely between junctions (see Figs.\ref{fig2} (a) and (c) vs. (b) and (d)).
Specifically, the spin-up conductances of the Ni-Ni/Ni(001) and Co-Co/Co(001)
junctions are almost the same (see Fig.\ref{fig2}(a)) while the spin-down ones
differ significantly (see Fig.\ref{fig2}(b)). For the Ni-X/Ni(001) junctions,
the same trend also exists (see Fig.\ref{fig2}(c) vs. (d)).
We note that the present result is consistent with our previous theoretical
finding: For SAMJs with similar atomic geometry but different species, their spin-down conductance is
sensitive to the species while the spin-up one is not\cite{xie}.
In Fig.\ref{fig2}, we can also find that for the
Ni-Fe/Ni(001) junction the spin-up conductance is greater than the spin-down
one, while for the other junctions the spin-down conductance becomes larger.

\begin{table*}[htb]
\caption{\label{tab:table1}%
Contact conductances and the two spin components (in
units of $e^2 /h$), as well as the SPR  of the five SAMJs in the FM and AFM
configurations, respectively. }
\begin{ruledtabular}
\begin{tabular}{ccccccc}
 &\scriptsize Co-Co/Co(001)& \scriptsize Ni-Ni/Ni(001)&
\scriptsize Ni-Fe/Ni(001)&
\scriptsize Ni-Co/Ni(001)&
\scriptsize Ni-Cu/Ni(001)\\
\hline
FM         &3.22    & 3.54 & 1.96 & 2.50 &2.66 \\
 $\uparrow$ & 1.38  & 0.97 & 1.17 & 1.04 &1.01 \\
 $\downarrow$ &1.84 & 2.57 & 0.79  & 1.46 &1.65 \\
 SPR    & -14.29\%	& -45.19\% &19.39\% &-17.13\% &-24.06\%\\
\hline
AFM         &3.26  & 3.44 & 2.12 & 2.64 &2.08 \\
 $\uparrow$ & 1.64  & 1.73 & 1.71 & 1.63 &1.22 \\
$\downarrow$ &1.62 & 1.71 & 0.41 & 1.01 &0.86 \\
 SPR    &0.61\%  &0.58\% &	61.32\% &	23.48\% &	17.31\%\\

\end{tabular}
\end{ruledtabular}
\end{table*}

\begin{figure}
\center
\includegraphics[width=8.5cm]{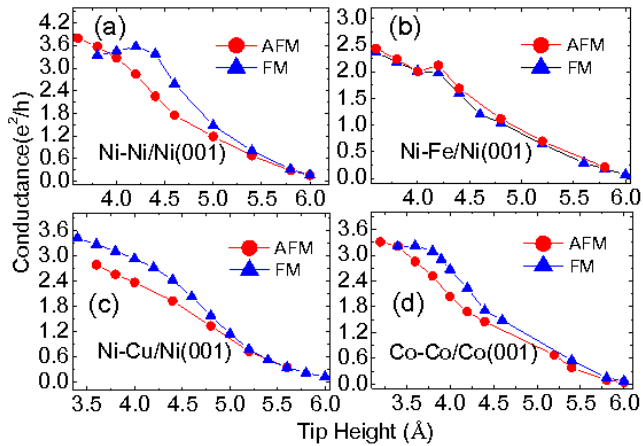}
\caption{Conductance as a function of the tip height for the SAMJs in the FM and
AFM configurations, respectively.}
\label{fig3}
\end{figure}

To investigate the influence of the spin configuration on conductance, we then
calculate the contact conductance for the AFM configuration, in which the
tip and right electrode are of the same spin alignment being opposite to
that of the adatom/surface and the left electrode.  The conductance as a
function of the tip height is plotted in Figs.\ref{fig3} and the values of the
determined contact conductance are listed in Table.\ref{tab:table1} together
with those for the FM configuration.
We find that, as compared
to the FM configuration, there is nearly no change in the total
conductance for the SAMJs with a magnetic adatom, i.e., the Co-Co/Co(001) and
Ni-X/Ni(001) (X=Fe, Co, Ni) junctions, as can be seen clearly in
Table.\ref{tab:table1} (The maximum change is only about 8\%).
In contrast, for the SAMJ with a non-magnetic adatom (Ni-Cu/Ni(001)), the change
in the contact conductance is as large as $22\%$ (see also Fig.\ref{fig3}(c)).

\begin{figure}
\center
\includegraphics[width=8.5cm]{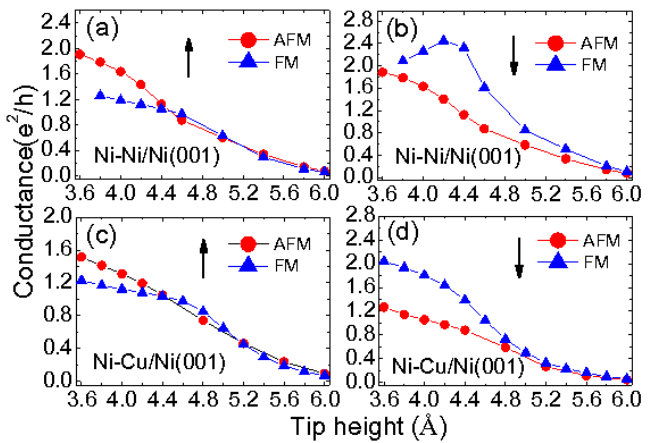}
\caption{Spin-dependent conductance as a function of the tip height for (a), (b) Ni-Ni/Ni(001) and (c), (d)
Ni-Cu/Ni(001) junctions in the FM and AFM configurations, respectively. The up
and down arrows indicate the spin-up and spin-down components, respectively. }
\label{fig4}
\end{figure}

To probe the origin of the different behavior for the two kinds of SAMJs (with a
magnetic or a non-magnetic adatom), we
study the two spin components of the total conductance, as plotted in
Fig.\ref{fig4} for the Ni-Ni/Ni(001) and Ni-Cu/Ni(001) junctions as examples.
We find that a general trend exists for all these SAMJs.
That is, when the spin configuration of the junctions changes from FM to AFM, the spin-up
conductance increases while the spin-down conductance decreases.
This trend can be clearly seen in Figs.\ref{fig4} (a) and (c) vs. (b) and (d).
For the junctions with a magnetic adatom, i.e., Co-Co/Co(001) and Ni-X/Ni(001) (X=Fe,
Co, Ni), the increase of the spin-up conductance counteracts the decease of the spin-down
component, thereby leading to a very small change in the total conductance and nearly no spin
valve effect. On the other hand, for the Ni-Cu/Ni(001) junction the increased spin-up conductance is less
than the decreased spin-down counterpart, thus resulting in a spin valve effect
with a variation of 22\% in the total conductance.

\begin{figure}
\center
\includegraphics[width=8.5cm]{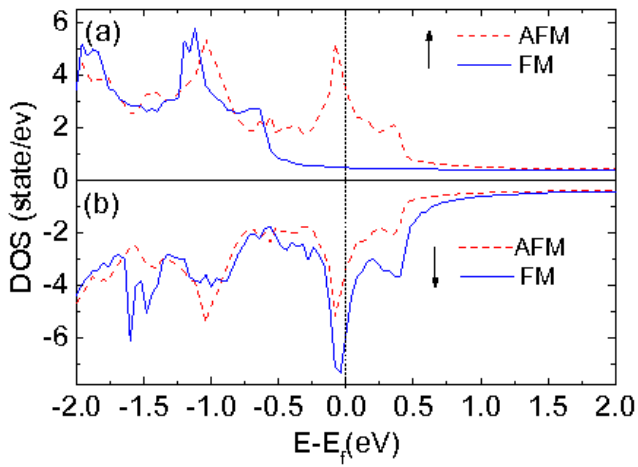}
\caption{(a) Spin-up and (b) spin-down PDOS of the Ni tip-apex atom and the Ni
adatom of the Ni-Ni/Ni(001) junction at the tip height of 3.8\AA\ for the FM and
AFM configuration, respectively.}
\label{fig5}
\end{figure}

To give a deeper understanding of this trend, we analyze the projected density
of states (PDOS) of the tip-apex atom and the adatom, which determines the
electron tunneling for a given tip height.
In Fig.\ref{fig5} we give the PDOS of the Ni adatom and
Ni tip-apex atom for the Ni-Ni/Ni(001) junction at the tip height of 3.8\AA.
The first thing to note is that for the AFM configuration the spin-up and
spin-down PDOS is basically symmetric as it should be because of the opposite
spin polarization of the two atoms, while for the FM
configuration the two PDOS become asymmetric, especially around the Fermi energy, due to
the same spin polarization of the two atoms.
As a result, when the junction changes from FM to AFM, the spin-up PDOS at the fermi energy is
largely increased (see Fig.\ref{fig5}(a)) while the spin-down PDOS is largely decreased (see Fig.\ref{fig5}(b)).
This explains the increase (decrease) of the spin-up (spin-down) conductance
when the spin configuration of the junction changes from FM to AFM.

We note that a spin-valve effect was found experimentally for a SAMJ
comprising a Cr-covered tip and a Co or Cr adatom on Fe nanoscale islands placed on a W(110)
substrate\cite{stm3}, where a variation of 40\% in the total conductact was
observed when the spin configuration changes from FM to AFM.
This is because the change of the spin-up conductance is much larger than
that of the spin-down one. In our cases, very weak spin-valve effects are obtained for the SAMJs with a magnetic
adatom due to the fact that the increase of the spin-up conductance counteracts the decrease
of the spin-down conductance when the junctions change from FM to AFM.

Finally, let us look at the spin filter effect and how it is influenced by the
change of the spin configuration of the SAMJs.
Tab. 1 lists the SPR of the five junctions in the FM and AFM configurations,
respectively. One can see that for the FM configuration the absolute value of
SPR ranges from 15\% to 45\%, showing a considerable spin filter effect.
Among the five junctions, only the Ni-Fe/Ni(001) junction has a
positive value of SPR while other junctions have negative ones.
This is because the spin-up channel dominates in the former while the spin-down
channel dominate in the latters.
When the junctions change from FM to AFM, the values of
SPR become all positive, since the spin-up conductance
increases and becomes larger than the spin-down one.
As a result, the absolute values of
SPR are reduced significantly because now the contributions from the spin-up and
spin-down channels become closer to each other except for the Ni-Fe/Ni(001) junction.
In the case of the Ni-Fe/Ni(001) junction, the total conductance for the FM
configuration already has the main
contribution from the spin-up channel, when the junction changes from FM to AFM its
spin-up conductance is further increased. Consequently, its SPR is enhanced from
19\% to 61\%.

\section{Conclusion}
In conclusion, by performing first-principles quantum transport calculations we have investigated the
spin-dependent conductance of five SAMJs in the FM and AFM spin
configurations, respectively. When the junctions change from FM to AFM, their spin-up
conductance increases while the spin-down one decreases. For the Co-Co/Co(001)
and Ni-X/Ni(001) (X=Fe, Co, Ni) junctions the increase of the spin-up conductance
counteracts with the decrease of the spin-down component, thereby showing no
spin valve effect.  For the Ni-Cu/Ni(001) junction, a spin valve effect is
observed with a variation of $22\%$ in the total conductance. In addition, the
spin filter effect of the Ni-Fe/Ni(001) junction is largely enhanced since the
spin-up electrons contribute more than the spin-down ones to the total
conductance. In contrast, the spin filter effect of the other junctions is
suppressed, as their conductance has the main contribution from the spin-down
electrons.
Our results show a general behavior of spin-dependent conductance for
these SAMJs under the change of the spin configuration, and the resulting
influences on the spin valve effect, as well as on the spin filter effect.

\begin{acknowledgments}
This work was supported by NSFC (Grant No. 51101102, No. 11174220 and No. 11004135), Leading
Academic Discipline Project of Shanghai Normal University (Grant
No. DZL712),   Innovation Researching Fund of Shanghai Municipal
Education Commission (Grant No. shsf020) and Program of Shanghai Normal University ( Grant No. DXL121),
as well as by  the MOST 973 Project under Grant No. 2011CB922204, and by King
Abdulaziz University (KAU) under Grant No. (49-3-1432/HiCi),
and the Unite Science Foundation of Guizhou Science and Technology Department (Grant No. QKHJZ-LKZS[2012]06).

\end{acknowledgments}


\end{document}